\documentstyle[prb,aps]{revtex}
\begin{document}
\draft
\title{
Square-lattice Heisenberg antiferromagnet with  two
kinds of \\
 nearest-neighbor regular bonds}

\author{N. B. Ivanov}
\address{Institute  for Solid  State
Physics, Bulgarian Academy of Sciences,\\
Tzarigradsko
chaussee-72, 1784 Sofia, Bulgaria}

\author{S. E. Kr\"uger and J. Richter}
\address{Institut f\"ur Theoretische Physik,
Universit\"at Magdeburg,\\
 P.O.Box 4120, D-39016 Magdeburg, Germany}
\date{\today}
\maketitle
\begin{abstract}
We study the zero-temperature phase diagram of
a square-lattice $S=1/2$ Heisenberg antiferromagnet with two types of
regularly distributed
nearest-neighbor exchange constants, $J_1>0$ (antiferromagnetic)
and $-\infty < J_2 <
\infty $,
using spin-wave series based on
appropriate mean-field Hamiltonian and exact-diagonalization data
for small clusters.
At a quasiclassical level, the model
displays two critical points separating the  N\'eel state
from (i) a helicoidal magnetic phase for relatively small
frustrating ferromagnetic couplings $J_2<0$ ($J_2/J_1<-1/3$
for classical spins), and (ii) a finite-gap
 quantum paramagnetic phase for
large enough antiferromagnetic exchange
constants $J_2>0$.
The quantum order-disorder transition (ii) is similar to the one
recently studied in two-layer Heisenberg antiferromagnets and is a
pure result of the zero-point spin fluctuations.
 On the other hand, the
melting of the N\'eel state  in the ferromagnetic region,
 $J_2<0$,
 is a combined effect of the
 frustration and
quantum spin fluctuations. The second-order
 spin-wave calculations of
the ground-state energy and on-site magnetization are in
 accord with  our
exact diagonalization data in a range away from the quantum
paramagnetic phase. In approaching the
phase boundary,
the theory fails due to the enhanced longitudinal
spin fluctuations, as it has recently been  argued by Chubukov
and Morr.
\end{abstract}


\section{Introduction}

The zero-temperature properties of a number of two-dimensional
 Heisenberg spin models displaying order-disorder quantum phase
transitions have recently attracted much attention. Most
efforts were connected with the
square-lattice Heisenberg antiferromagnets
 with frustrating second $J_2$  and third
$J_3$  nearest-neighbor bonds \cite{j1j2}
as well as with
the two-layer quantum Heisenberg
models \cite{twolayer}.

The mentioned models  present  concrete
realizations of the quantum order-disorder
transition considered by Chakravarty, Halperin, and Nelson
\cite{chakravarty} using
a macroscopical approach based on the quantum
non-linear $\sigma$ model
in $2+1$ dimensions. Notice that in the $J_1-J_2-J_3$
models the destruction of the  N\'eel state is a result of both
the frustration and zero-point spin fluctuations,
 whereas the two-layer
antiferromagnet can be turned through an order-disorder transition
by varying the antiferromagnetic exchange coupling
 between the planes. The latter model
constitutes an ideal system where the zero-point spin fluctuations
 destroy the classical N\'eel state.

Less studied are the $2D$ Heisenberg antiferromagnets
 with random ferromagnetic
exchange bonds \cite{ferbonds} which are currently
considered as an appropriate  model of
the copper-oxygen sheets of lightly doped insulating high-$T_c$
superconductors. The idea comes from Aharony and co-workers
\cite{aharony} who
suggested that the holes introduced into the $\rm CuO$ layers are
localized at individual oxygen sites. The coupling between the copper
and oxygen spins results in an effective ferromagnetic coupling
between the copper spins situated near an oxygen.

In this paper we study a quantum
 Heisenberg antiferromagnet
with two kinds of nearest-neighbor exchange
couplings, regularly distributed on
the square lattice, Fig.1. The relevance of the model is
twofold. First, it combines the features of
 the  $J_1-J_2-J_3$ and two-layer models in displaying
the mentioned two mechanisms of destruction of the N\'eel phase
and, on the other hand, the model can be considered
as a prototype of some more
realistic  antiferromagnetic models containing ferromagnetic bonds.

 The Hamiltonian of the system reads
\begin{equation}
H=J_1H_1+J_2 H_2 \equiv J_1\sum_{<ij>}{\bf S_iS_j}+
J_2\sum_{(ij)}{\bf S_iS_j},
\label{ham}
\end{equation}
where the sums run over the nearest-neighbor $J_1$ and $J_2$ bonds,
respectively. In what follows we frequently use the notations
 $\alpha \equiv J_2/J_1, J_1 \equiv 1$.
 In the antiferromagnetic region $J_2>0$, Singh, Gelfand, and
Huse \cite{singh} have found for Eq.(\ref{ham}) a critical point
$\alpha_s\approx 2.56$ separating
 the classical N\'eel ordered and a finite-gap quantum paramagnetic
  phases, using series
expansions around the dimer
Hamiltonian $H_2$,. They also argued that the
critical exponents are consistent with those of the $3D$ classical
Heisenberg model. In this study we analyze the zero-temperature
phase diagram of the model in the
range $-\infty < J_2 <
\infty $ both analytically, reducing the problem to an interacting
Bose gas,
 and numerically, throughout exact
diagonalization of small clusters.
For  ferromagnetic exchange bonds, at $\alpha =-1/3$ for classical
spins,
we found
 another critical point,
 dividing the N\'eel phase
from a helicoidal magnetic phase, Fig.1. The
latter phase reduces to a three-sublattice N\'eel
state in the limit $J_2=- \infty$. The classical
helicoidal phase is
described
by the equations
\begin{eqnarray}
{\bf S_A(R)}&=&{\bf \hat{u}}\cos{\bf QR}+{\bf \hat{v}}\sin{\bf QR},
\nonumber\\
{\bf S_B(R+\hat{x})}&=&{\bf \hat{u}}\cos({\bf QR}+\pi+3\Phi)
+{\bf \hat{v}}\sin({\bf QR}+\pi+3 \Phi),
\end{eqnarray}
where the classical spins ${\bf S_A}$ and ${\bf S_B}$ belong
 to the sublattice $A$ and
$B$, respectively. ${\bf \hat{u}}$
 and ${\bf \hat{v}}$ are perpendicular
unit vectors in the spin space; ${\bf R}$ runs over the sites of
 the sublattice $A$.
 The pitch vector $\bf Q$ and the phase
$\Phi$ are determined through the relations
\begin{eqnarray}
\cos Q_{x(y)}&=&\frac{1}{2}\left(\frac{1}{|\alpha |}-1\right),
\hspace{1cm}
Q_{y(x)}=0,\\
\Phi&=&\frac{Q_{x(y)}}{2}, \hspace{1cm}
\alpha \le -\frac{1}{3}.\nonumber
\end{eqnarray}
The angle $\Phi$ can be chosen as an order-parameter of the
continuous classical transition.

 Below we concentrate on an analysis
of the role of quantum spin fluctuations in
the N\'eel phase of Eq.(\ref{ham}). To this
end, in the next section we follow a method overcoming
 the problems of the standard spin-wave theory (SWT) connected to
logarithmic divergences at the classical transition points. Then
we construct spin-wave series for the ground-state energy
and on-site magnetization, and discuss the changes in the phase
diagram  implied by the interaction.
 In the last section we present exact diagonalization (ED)
data for small clusters and compare them with the theory.

\section{Spin-wave series for the ground-state energy
and on-site magnetization}

We  proceed with a short discussion of the results implied by the
linear  spin-wave
 theory (LSWT).
 LSWT predicts the existence of a region around
the classical  transition  point $\alpha_c=-1/3$,
where the  magnetic states are completely melt by the zero-point
spin fluctuations, Fig.2. In this approximation,
 the spin-liquid phase exists
for arbitrary $S$ due to the logarithmic
divergence at $\alpha=-1/3$ in the integral
\begin{equation}
  \delta m= \frac{1-\kappa_0}{N}\sum_{\bf k}\frac{1}
  {\varepsilon_{s \bf
  k}}-\frac{1}{2} \sim \ln\frac{1}{|\alpha-\alpha_c|},
\label{m1}
\end{equation}
defining the reduction of the on-site magnetization
 $\delta m=S-<S_z>$.
Here $N$ is the number of lattice sites, $\kappa_0=(1-\alpha)/4$.
 This behavior of $\delta m$
  can be regarded as an aftereffect of the classical "soft lines"
  at the critical point in the low-energy magnon dispersion function
\begin{equation}
   \varepsilon_{s \bf k} \sim \sqrt{c_1(\alpha-\alpha_c)k^2_x
   +c_2k^2_y}, \hspace{1cm} |{\bf k}|\approx 0,\pi.
\end{equation}
Due to the continuous character of the classical
 transition, one could
 expect the same behavior when  the interaction is switched on.
 A consideration valid in the large-$\rm S$ limit,
 and based on an appropriate quantum nonlinear $\sigma$ model
 on the critical line \cite{ioffe}, predicts the
 existence of a disordered spin-liquid phase
 in an exponentially small region  around the critical
 line.  There are some indications that the
 spin-liquid phase survives for arbitrary $\rm S$  up to the
 extreme quantum limit $\rm S=1/2$ \cite{ferrer}.
 In our case, however, the structure of the quasiparticle
 spectrum (i.e., the number of Goldstone modes) is changed, so that
 the problem requires a special study \cite{chub1} of the
 helicoidal phase, wich
is beyond the scope of the present paper.

LSWT also qualitatively predicts the transition to
a finite-gap quantum paramagnetic state
in the antiferromagnetic, $J_2>0$, region of the phase diagram
\cite{richter1}.
 However,
the estimate  $\alpha_s \approx
10.6$  considerably exceeds the series estimate $\alpha_s\approx
2.56$.
 The
situation is analogous to the one in  two-layer antiferromagnets
\cite{chub2}, i.e., SWT is not successful because of
 the enhanced longitudinal spin fluctuations.  One of the
 purposes of the present study, using also exact diagonalization
 data, is to see to what extent SWT can be trust in approaching
 the quantum paramagnetic phase. Our model has the advantage
of permitting exact
 diagonalization studies for large enough clusters.

\subsection{The method}

As mentioned above, the standard spin-wave series
are divergent at  the classical transition point $\alpha_c=-1/3$.
For example, the first-order correction to  the
 on-site magnetization $\Delta m^{(1)}$ is expressed through the
 following divergent integrals
 \begin{eqnarray}
\Delta m^{(1)}&=& \frac{a_0}{S} \frac{1}{N}\sum_{\bf k}
\frac{(\nu_{\bf k}-\kappa_0
\cos k_x)(\cos k_x-\nu_{\bf k})-\kappa_0 \sin^2k_x}
{\varepsilon_{s\bf k}^3},\\
           a_0&=&-\frac{\kappa_0}{N}
\sum_{\bf k}
\frac{\kappa_0 (1-\nu_{\bf k}\cos k_x )+\nu_{\bf k}(\nu_{\bf k}-
\cos k_x)}
{\varepsilon_{s \bf k}}, \nonumber
\end{eqnarray}
 where $\nu_{\bf k}=(\cos k_x+\cos k_y)/2$, $\varepsilon_{s \bf k}
 =\sqrt{(1-\nu_{\bf k}^2)-2\kappa_0
(1-\nu_{\bf k}\cos k_x)}$.
Since in the SWT formalism the on-site magnetization is a two-boson
operator, it is clear that the divergences in the first-order
correction $\Delta m^{(1)}$ come from two-boson interaction
terms. A straightforward way to improve the situation
 is to use  a
normal-ordered operator form for the interaction.
 This approach leads to renormalization
of the quasiparticle excitation spectrum and to
some resumed SWT series starting now from
an appropriate mean-field theory (MFT).
 Instead of $1/S$, the new series
are, however,  governed by a formal small
 parameter $\lambda$ introduced
 in front of the quartic
Dyson-Maleev normal-ordered boson interaction.
Therefore, the quality of
the series should be controlled through another method.
In our study
we present exact diagonalization  data for small clusters.

Let us briefly describe the procedure leading to the spin-wave
series.
SWT requires three subsequent transformations.
 Through a standard  Dyson-Maleev bosonization scheme,
  the original spin Hamiltonian transforms
into  equivalent  boson Hamiltonian. In the usual SWT
 the two-boson terms scaling with $S$
constitute the free-boson
Hamiltonian $H_0$ which fixes the $1/S$ series.
As it was mentioned above, in our case
more convenient is to diagonalize
the quadratic part of the Hamiltonian by means of a
generalized Bogoliubov transformation, also touching on the
anharmonic terms \cite{rastelli,kopietz,canali1}.
 The new quadratic Hamiltonian
$H_0$ can be obtained in the following way. First, we
subsequently do the Fourier
and  quasiparticle Bogoliubov transformations, the latter being
defined through the relations
\begin{equation}
a_{\bf k}=u_{\bf k}\alpha_{\bf k}+v_{\bf k} \beta_{\bf
k}^{\dag},\hspace{1cm}
b_{\bf k}=u_{\bf k}\beta_{\bf k}+v_{\bf k}
\alpha_{\bf k}^{\dag}, \hspace{1cm}
|u_{\bf k}|^2-|v_{\bf k}|^2=1.
\end{equation}
In a quasiparticle representation,
the boson Hamiltonian takes the form
\begin{equation}
H=E_0+H_0+ V_{DM},
\label{ham2}
\end{equation}
where the Dyson-Maleev interaction $V_{DM}$ is in a
normal-ordered operator form (see Appendix A). Then  $E_0$
is the mean-field
ground-state energy and $H_0$ is a quadratic free-particle
 Hamiltonian, containing also contributions from the operator ordering
of the quartic interaction term.

 The diagonalization of
$H_0$, which in terms of quasiparticle operators  reads
\begin{equation}
H_0=\sum_{\bf k}\Big[ E_{\bf k}(\alpha^{\dag}_{\bf k}\alpha_{\bf k}
+\beta^{\dag}_{\bf k}\beta_{\bf k})+
(B_{\bf k}\alpha^{\dag}_{\bf k}\beta^{\dag}_{\bf k}+h.c.)\Big] ,
\label{hoo}
\end{equation}
 requires
 $B_{\bf k}=0$.
In principle, the condition $B_{\bf k}=0$ leads to some
 integral equations for $u_{\bf k}$ and $v_{\bf k}$ studied
in the general case by Obukhov \cite{obukhov}.
Recently, similar method has been
applied to the $J_1-J_2$ model \cite{gochev}.

Note that the so-defined mean-field theory can also be
regarded as a result of the variational equations
\begin{equation}
 \frac{\delta}{\delta u_{\bf k}}\frac{<\Psi |H|\Psi >}
{<\Psi |\Psi >}=0, \hspace{1cm}
\frac{\delta}{\delta v_{\bf k}}\frac{<\Psi |H|\Psi >}
{<\Psi |\Psi >}=0
\end{equation}
(under the constraint $|u_{\bf k}|^2-|v_{\bf k}|^2=1$),
defining the generalized Bogoliubov transformation.
The quality of the discussed perturbation scheme mainly depends on
the quasiparticle interaction strength playing the
role of a physical small parameter. Particularly,
the latter is expected to
increase near the  critical points.
The model Hamiltonian (\ref{ham}) represents a convenient
system to check the theory due to the existing
two critical points and a numerical-series estimate \cite{singh}
for the  order-disorder transition point.

\subsection{Zeroth-order mean-field approximation}

The above program is straightforward for Eq.(\ref{ham}) and
leads to a renormalization of the bare coupling $(1-J_2/J_1)
\longrightarrow (1-J_2/J_1)Z$. The implicit equation for
the factor $Z=Z(S,\alpha)$ reads
\begin{equation}
Z=\frac{S-a_1}{S-a_2},
\label{z}
\end{equation}
where
\begin{eqnarray}
                a_1&=&-\frac{1}{2}+\frac{1}{N}
                \sum_{\bf k}\frac{1-\nu_{\bf k}
\cos k_x}{\varepsilon_{\bf k}},\hspace{1cm}
a_2=-\frac{1}{2}(1-\kappa)+\kappa a_1+\frac{1}{N}
\sum_{\bf k}\varepsilon_{\bf k},\\
\varepsilon_{\bf k}&=&\sqrt{(1-\nu_{\bf k}^2)-2\kappa
 (1-\nu_{\bf k} \cos k_x)},
\hspace{1cm} \kappa=(1-\alpha)\frac{Z}{4}.
\end{eqnarray}
 The renormalization factor $Z$ defines (up to an irrelevant
phase factor) the Bogoliubov coefficients $u_{\bf k}$ and
$x_{\bf k}=- v_{\bf k}/u_{\bf k}$
\begin{eqnarray}
     u_{\bf k}&=&\sqrt{ \frac{1-\kappa+\varepsilon_{\bf k}}
{2\varepsilon_{\bf k}}} \nonumber\\
 \Re(x_{\bf k})&=&\frac{\nu_{\bf k}-\kappa \cos k_x}
 {(1-\kappa+\varepsilon_{\bf k})},
\label{uv}\\
 \Im(x_{\bf k})&=&\frac{-\kappa \sin k_x}
 {(1-\kappa+\varepsilon_{\bf k})}, \nonumber
\end{eqnarray}
Notice that one cannot choose the phase factor so as to
make both  $u_{\bf k}$ and $v_{\bf k}$
real functions.

The quasiparticle excitation spectrum  reads
\begin{equation}
E_{\bf k}=4S\left( 1+\frac{r}{2S}\right)
\varepsilon_{\bf k},
\label{ek}
\end{equation}
where $r$ is a numerical factor analogous to  Oguchi's
correction \cite{oguchi}, $r=-2a_2$.

The ground-state energy $E_0$ and on-site magnetization $m_0$ in
this approximation are
\begin{equation}
\frac{E_0}{2N}=-(1-\kappa Z)( S-a_2 ) ^2,
\label{eee}
\end{equation}
\begin{equation}
m_0=S+\frac{1}{2}-\frac{1-\kappa}{N}\sum_{\bf k}
\frac{1}{\varepsilon_{\bf k}}.
\label{mmm}
\end{equation}
We will see below that the mean-field estimate
 for the ground-state energy
$E_0$ is in agreement with ED  in
a large region of the phase diagram. The on-site
magnetization $m_0$ vanishes at $\alpha_c\approx -3.0$,
 i.e., there is a
shift of the classical transition point
   towards the helicoidal
 phase. This phenomenon
  is known as "Villain's
 order from disorder"\cite{villain}.
 The transition to  quantum paramagnetic state in MFT  is
 located at
 $\alpha_s\approx 5.7$, which is approximately twice larger than
 the series estimate. Below we
 show that the second-order correction to this result improves the
 estimate.

\subsection{Second-order corrections to $E_0$ and $m_0$}
Now we proceed with a calculation of the second-order $O(V_{DM}^2)$
 corrections
 to
$E_0$ and $m_0$.
 It is easy to see that the first-order corrections
  both  to $E_0$ and $m_0$
 identically vanish for the normal-ordered interaction $V_{DM}$.
 The second-order correction to $E_0$
  has the standard form (Appendix B)
 \begin{equation}
 \frac{\Delta E^{(2)}}{2N}=-\left( \frac{2}{N}\right)^3
 \sum_{1-4}\Delta (1+2-3-4)
 \frac{V^{(7)}(1234)V^{(8)}(3412)}{E_1+E_2+E_3+E_4},
\end{equation}
where $E_{\bf k}$ are the quasiparticle energies Eq.(\ref{ek}).
The expression for the second-order correction
to the on-site magnetization consists of
two parts (see Appendix B)
\begin{equation}
 \Delta m^{(2)}=\Delta m^{(2)}_a +\Delta m^{(2)}_b,
\end{equation}
where
\begin{eqnarray}
\Delta m_a^{(2)} &=& -\frac{64S}{N^3}(1+\frac{r}{2S})(1-\kappa
)\sum_{1-4}\frac{\Delta(1+2-3-4)}{(E_1+E_2+E_3+E_4)^2} \nonumber \\
                 & & \times V^{(7)}(1234)V^{(8)}(3412)\left(
\frac{1}{E_1}+\frac{1}{E_2}+\frac{1}{E_3}+
\frac{1}{E_4}\right),
\label{ma}
\end{eqnarray}

\begin{equation}
\Delta m_b^{(2)}= \frac{16}{N^3}
\sum_{1-4}\frac{\Delta({
1+2-3-4})}{E_1+E_2+E_3+E_4}W(1234),
\label{mb}
\end{equation}
\begin{eqnarray}
W(1234)&=&V^{(7)}(1234)\left[ \frac{u_2^2x_2^*}{E_2}V^{(5)}(3412)+
\frac{u_3^2x_3^*}{E_3}V^{(2)}(3412)\right] \nonumber \\
       &+&V^{(8)}(3412)\left[ \frac{u_1^2x_1}{E_1}V^{(6)}(1234)+
\frac{u_4^2x_4}{E_4}V^{(3)}(1234)\right].
\end{eqnarray}

In the last equations the sums run over
the vectors  $\bf k_1,k_2,k_3,k_4$ from
 the magnetic Brillouin zone; $\Delta({\bf k})$ is the Kronecker
function. The explicit expressions for the vertex functions
$V^{(i)}\equiv V^{(i)}({\bf k_1k_2,k_3k_4})$, $i=1,...,9,$
are given in Appendix A.
A straightforward task is to show that the above $6D$ integrals
defining $\Delta E^{(2)}$ and $\Delta m^{(2)}$ are
convergent, using the vertex expansions near ${\bf k}=0$ and
${\bf k}=\pi$
\cite{kopietz}.

The $6D$ integrals written above were calculated using a standard
Gaussian method for a set of lattices with up to $M=(20\times
20)^3$ cells (each containing 4 points) and then the results
 were extrapolated to $M=\infty$.
 Starting from the $(4\times 4)^3$-cell  lattice the
results change monotonously and ensure the precision presented in
Tables 1, 2.

\section{Comparison with exact-diagonalization data and discussion}

For the pure $J_1=J_2$ Heisenberg model, the series for
the ground-state energy $E$ and on-site magnetization $m$ read
\begin{equation}
E/2N=-0.33521+0.00019\lambda^2+O(\lambda^3)= -0.33502+O(\lambda^3),
\end{equation}
\begin{equation}
m=0.3034+0.0022\lambda^2+O(\lambda^3)=0.3056+O(\lambda^3),
\end{equation}
where $\lambda=1$. The estimates for $E$ and $m$ are
in accord with previous large$-S$ \cite{canali1,hamer} and
numerical \cite{singh2} series results.
The above series show that in the pure model the second-order
$O(V_{DM}^2)$ corrections are numerically small ( i.e.,
 the effective quasiparticle interaction
is weak) which means that the theory is self-consistent up to this
order. As a matter
of fact, the corrections $\Delta E^{(2)}$ do not exceed $3.5\%$ of
the MFT values
in the whole range $-1.2<\alpha <5.0$ where the on-site
magnetization $m$ is positive. Notice that
the second-order correction $\Delta E^{(2)}$ becomes negative
\cite{ivanov2}
 for $\alpha < -0.8$ and
$\alpha > 1.4$, Table 1. The second-order correction to the on-site
magnetization $\Delta m^{(2)}$ is
less than $10\%$ as compared to $m_0$ in the
 interval $-0.9 < \alpha < 3.4$.
$\Delta m^{(2)}$ is negative for $\alpha <-0.8$ and $\alpha > 1.6$,
Table 2.

In Fig.3 we show the SWT results for the ground-state energy
of Eq.(\ref{ham}) compared with ED data .
The ED data are slightly size dependent, so that, for sake of
clarity, we present only
 the results for a cluster with $N=26$ spins.
 In the parameter range $-0.4< \alpha < 2.0$
we find
excellent agreement with ED data.
 Remember that $\alpha_s\approx
2.56$ is the numerical-series estimate for the order-disorder
critical point \cite{singh}, and $\alpha=-1/3$ is the LSWT
estimate for the second critical point of Eq.(\ref{ham}).
For $\alpha >2.0$
 and $\alpha < -0.4$ one observes
an increasing discrepancy with  ED data. At the same time, the
theory implies very small
corrections to the zeroth-order energy $E_0$ in the whole range
$-1.2 < \alpha < 4.0$.
 Analogous  behavior of the
SWT series has recently been indicated by Chubukov and Morr
\cite{chub2} in the two-layer
Heisenberg  antiferromagnet when approaching the quantum
paramagnetic phase. It was argued, using an
appropriate bosonization scheme \cite{chub3}, that
the spin-wave description fails due to the enhanced
 longitudinal spin fluctuations near the quantum
paramagnetic phase. Near
the classical transition point $\alpha_c=-1/3$, where the melting of
the N\'eel
state is a result both of the frustration and zero-point
spin fluctuations, the discussed discrepancy is
less pronounced.

As a whole, the  second-order correction to the on-site
 magnetization $\Delta m^{(2)}$, Fig. 4,
improves the estimates for the helicoidal $\alpha_c\approx -1.2$ and
paramagnetic $\alpha_s\approx 5.0$ critical points. However,
the estimate for $\alpha_s$
is still twice larger than the numerical-series result $\alpha_s
\approx 2.56$. At the same time,
 for instance at $ \alpha=2.5$, the  correction
$\Delta m^{(2)}$ is less
than $3\%$ from the mean-field value $m_0$.
In Fig. 4 we also present extrapolated ED data  for the
on-site magnetization deduced from the antiferromagnetic
structure factor
$M_s^2=\frac{1}{N^2}\sum_{i,j}(-1)^{i+j}<{\bf S_iS_j}>$
 using
the cluster set $N=16, 18, 20, 26$ and the extrapolation
ansatz $M_s^2=M_s^2(\infty ) + constN^{-\frac{1}{2}}$.
The ED estimate $m \approx 0.28$ at $\alpha=1$
is smaller (but in agreement with other ED data using
also the $N=36$ cluster \cite{schulz})
than the SWT and numerical-series results
mentioned above. The above
discrepancy probably indicates the importance of the next-order term
in the extrapolation ansatz \cite{bernu}.
Notice that our ED estimate
$\alpha_s\approx 2.4$  is in  reasonable agreement with
 the numerical-series
result $\alpha_s\approx 2.56$.

 The SWT result  $\alpha_c \approx -1.2$, Fig. 4,
 also exceeds the estimate  obtained through an exact
  diagonalization of small
clusters with $N=16, 18, 20$ and $26$ spins.
 The ED data predict a phase-transition point $\alpha_c$
  which is close to the
classical result $\alpha_c=-1/3$. Here the situation is analogous to
the one in the $J_1-J_2$ model containing frustrating
diagonal bonds. For this model, both LSWT and
the recent ED data\cite{schulz,richter2,einarsson} predict a
complete melting of the N\'eel phase at $J_2/J_1
\approx 0.4$ , whereas an inclusion of the spin-wave
interaction
 changes the estimate at least up to  the  classical
transition point $J_2/J_1=0.5$\cite{gochev,ivanov}.
 There is a smaller SWT
estimate $J_2/J_1\approx 0.43$ based
 on the standard large-S expansion
\cite{igarashi2}, however it comes from large
negative corrections $\Delta m^{(2)}$  reducing the effect
of the diverging
 $\Delta m^{(1)}$ term, Eq.(\ref{m1}).

In conclusion, we have studied the influence of quantum spin
fluctuations on the zero-temperature phase
diagram of a $S=1/2$ square-lattice Heisenberg antiferromagnet
with two kinds of nearest-neighbor exchange bonds.
The phase diagram is found to
contain a helicoidal magnetic phase and a finite-gap
quantum paramagnetic phase, both of them being separated
through continuous transitions
 from the classical N\'eel phase. We constructed  an appropriate
 spin-wave
 expansion and  used exact-diagonalization data to find the changes
of the phase diagram implied by the zero-point spin fluctuations.
The latter approach reveals in particular
the limits of the spin-wave description coming from the
enhanced longitudinal spin fluctuations
near the phase boundaries.  The disagreement with ED data is
 strongly
revealed in approaching the quantum-paramagnetic-phase boundary.
In a forthcoming paper \cite{richter1} we present, in particular,
ED-based
analysis concerning the finite-gap quantum paramagnetic phase.

\acknowledgments

 N.I. was supported by the
National Science Foundation, Grant $\Phi412/94$. S.K. and J.R.
were supported by DFG, Grant Ri-615/1-2.
N.I. acknowledges the kind hospitality
of the University at Magdeburg where part of this work
was done. The visit in uni-Magdeburg was supported by DFG.


\appendix

\section{Dyson-Maleev Vertices}
 The normal-ordered Dyson-Maleev interaction  $V_{DM}$
  can be written in
 the following  form \cite{canali}
\begin{eqnarray}
V_{DM} &=&- \frac{2}{N} \sum_{1-4} \Delta (1+2-3-4) \nonumber \\
&\times& \Big[
V^{(1)} \alpha^{\dag}_1 \alpha^{\dag}_2 \alpha_3 \alpha_4 +
V^{(2)} \alpha^{\dag}_1 \beta_2 \alpha_3 \alpha_4+
V^{(3)} \alpha^{\dag}_1 \alpha^{\dag}_2 \beta^{\dag}_3 \alpha_4
+ V^{(4)} \alpha^{\dag}_1 \alpha_3 \beta^{\dag}_4 \beta_2 +
V^{(5)} \beta^{\dag}_4 \alpha_3 \beta_2 \beta_1 \nonumber \\
&+&V^{(6)} \beta^{\dag}_4 \beta^{\dag}_3 \alpha^{\dag}_2
\beta_1
 + V^{(7)} \alpha^{\dag}_1 \alpha^{\dag}_2 \beta^{\dag}_3
\beta^{\dag}_4 +
V^{(8)} \beta_1 \beta_2 \alpha_3 \alpha_4+
V^{(9)} \beta^{\dag}_4 \beta^{\dag}_3 \beta_2 \beta_1
\Big] ,
\label{vdm}
\end{eqnarray}
where the sum runs over the vectors  $\bf k_1,k_2,k_3,k_4$
in the magnetic Brillouin zone and the Kronecker function
$\Delta({\bf k})$ expresses the conservation of momentum
to within a reciprocal-lattice vector ${\bf G}$.
For the Hamiltonian (\ref{ham}) the nine complex vertex functions are

\begin{eqnarray}
V^{(1)}(1234) &=& {\cal U}(1234)
\Big( x_1\left[ x_4^{\ast}(\gamma_{1-4}
-x_3^{\ast}\gamma_{1-3-4})-(\gamma_1-x_3^{\ast}\gamma_{1-3})
\right] \nonumber \\
              &+& x_2\left[ x_4^{\ast}(\gamma_{2-4}
-x_3^{\ast}\gamma_{2-3-4})-(\gamma_2-x_3^{\ast}
\gamma_{2-3})\right]\Big),\\
V^{(2)}(1234) &=& 2{\cal U}(1234)
\Big( x_1x_2^{\ast}\left[ x_4^{\ast}
(x_3^{\ast}\gamma_{1-3-4}
-\gamma_{1-4})-(x_3^{\ast}\gamma_{1-3}-
\gamma_{1})\right] \nonumber \\
              &+& \left[ x_4^{\ast}(x_3^{\ast}\gamma_{2-3-4}
-\gamma_{2-4})-(x_3^{\ast}\gamma_{2-3}-\gamma_{2})\right] \Big),\\
V^{(3)}(1234) &=& 2{\cal U}(1234)
\Big( x_1\left[ x_4^{\ast}(\gamma_{1-3-4}
-x_3\gamma_{1-4})-(\gamma_{1-3}-x_3\gamma_{1})\right] \nonumber \\
              &+& x_2\left[ x_4^{\ast}(\gamma_{2-3-4}
-x_3\gamma_{2-4})-(\gamma_{2-3}-x_3\gamma_{2})\right]  \Big),\\
V^{(4)}(1234) &=& 4{\cal U}(1234)\Big( x_1x_2^{\ast}\left[ x_4
(x_3^{\ast}\gamma_{1-3}
-\gamma_{1})-(x_3^{\ast}\gamma_{1-3-4}-
\gamma_{1-4})\right]\nonumber \\
              &+& \left[ x_4(x_3^{\ast}\gamma_{2-4}
-\gamma_{2})-(x_3^{\ast}\gamma_{2-3-4}-
\gamma_{2-4})\right] \Big) ,\\
V^{(5)}(1234) &=& 2{\cal U}(1234)
\Big( x_2^{\ast}\left[ (x_3^{\ast}
\gamma_{1-3-4}
-\gamma_{1-4})-x_4(x_3^{\ast}\gamma_{1-3}
-\gamma_{1})\right]\nonumber \\
              &+& x_1^{\ast} \left[ (x_3^{\ast}\gamma_{2-3-4}
-\gamma_{2-4})-x_4(x_3^{\ast}\gamma_{2-3}-
\gamma_{2})\right] \Big) ,\\
V^{(6)}(1234) &=& 2{\cal U}(1234)\Big( x_1^{\ast}x_2\left[ x_4
(x_3\gamma_{2}
-\gamma_{2-3})-(x_3\gamma_{2-4}-\gamma_{2-3-4})\right]\nonumber \\
              &+& \left[ x_4(x_3\gamma_{1}
-\gamma_{1-3})-(x_3\gamma_{1-4}-\gamma_{1-3-4})\right] \Big) ,\\
V^{(7)}(1234) &=& {\cal U}(1234)\Big( x_1\left[ x_4(\gamma_{1-3}
-x_3\gamma_{1})-(\gamma_{1-3-4}-x_3\gamma_{1-4})
\right] \nonumber \\
              &+& x_2\left[ x_4(\gamma_{2-3}
-x_3\gamma_{2})-(\gamma_{2-3-4}-x_3\gamma_{2-4})\right]\Big),\\
V^{(8)}(1234) &=& {\cal U}(1234)\Big( x_2^{\ast}\left[ (x_3^{\ast}
\gamma_{1-3}
-\gamma_{1})-x_4^{\ast}(x_3^{\ast}\gamma_{1-3-4}-\gamma_{1-4})
\right]\nonumber \\
              &+& x_1^{\ast} \left[ (x_3^{\ast}\gamma_{2-3}
-\gamma_{2})-x_4^{\ast}(x_3^{\ast}\gamma_{2-3-4}-\gamma_{2-4})
\right] \Big) ,\\
V^{(9)}(1234) &=& {\cal U}(1234)
\Big( x_2^{\ast}\left[ x_4(\gamma_{1-3}
-x_3\gamma_{1})-(\gamma_{1-3-4}-x_3\gamma_{1-4})
\right] \nonumber \\
              &+& x_1^{\ast}\left[ x_4(\gamma_{2-3}
-x_3\gamma_{2})-(\gamma_{2-3-4}-x_3\gamma_{2-4})\right]\Big),
\end{eqnarray}
where ${\cal U}(1234)\equiv u_1u_2u_3u_4$
and the Bogoliubov coefficients $u_{\bf k}$ and $x_{\bf k}\equiv -
v_{\bf k}/u_{\bf k}$ are dedined by Eq.(\ref{uv}).
 $\gamma_{\bf k}$
is a complex lattice structure factor given by
\begin{equation}
\gamma_{\bf k}=\nu_{\bf k}-\kappa_0\cos k_x+i\kappa_0\sin k_x,
\end{equation}
where $\nu_{\bf k}=(\cos k_x+\cos k_y)/2$,
$\kappa_0\equiv (1-\alpha )/4$.
In the symmetric $J_1=J_2$ model, the vertex functions are real
and reduce to the well-known Dyson-Maleev vertices \cite{canali}.

\section{Calculation of $\Delta  m^{(2)}$ }

A straightforward way to find the second-order correction to
 on-site magnetization $\Delta m^{(2)}$ is to introduce in $H$
a perturbing staggered magnetic field $h$ (see, e.g.,
Ref.\onlinecite{gochev}).
 The perturbed Hamiltonian
 reads
 \begin{equation}
 H(h)=H-hN\hat{m}\equiv E_0(h)+H_0(h)+V_{DM},
 \label{hh}
 \end{equation}
 where the on-site magnetization operator has the form
 \begin{equation}
\hat{m}=S-\frac{2}{N}\sum_{\bf k}|v_{\bf k}|^2-\frac{1-\kappa}{N}
\sum_{\bf k}\frac{1}{\varepsilon_{\bf k}}\left[
(\alpha^{\dag}_{\bf k}\alpha_{\bf k}+
\beta^{\dag}_{\bf k}\beta_{\bf k})
-2u^2_{\bf k}
(x_{\bf k}\alpha^{\dag}_{\bf k}\beta^{\dag}_{\bf k}+h.c.)\right].
\end{equation}
In the new basis of $H_0(h)$ the second-order correction to
the ground-state energy reads
\begin{equation}
\Delta E^{(2)}(h)=-\sum_{[\rho_h]}\frac{
<0_h|V_{DM}|\rho_h><\rho_h|V_{DM}|0_h>}{E_{\rho_h}},
\label{eh}
\end{equation}
where $|\rho_h>$ are four-particle eigenstates of  $H_0(h)$.

 $\Delta m^{(2)}$ can be obtained by  applying the Hellmann-Feynman
 theorem to Eq.(\ref{hh}) and then using Eq.(\ref{eh})
\begin{equation}
\Delta m^{(2)}=-\frac{1}{N}\frac{\partial}{\partial h}\Delta
E^{(2)}(h)_{h=0}.
\label{mag}
\end{equation}
There are two contributions to Eq.(\ref{mag})
$\Delta m^{(2)}_a$ and $\Delta m^{(2)}_b$,
 coming respectively from the
$O(h)$ terms in the dominator and numerator of Eq.(\ref{eh}).
A simple calculation using
\begin{equation}
E_{\bf k}(h)=E_{\bf k}+(1-\kappa)
\frac{h}{\varepsilon_{\bf k}}+O(h^2),
\end{equation}
gives the correction $\Delta m^{(2)}_a$, Eq.(\ref{ma}).

The second contribution $\Delta m^{(2)}_b$ results from the $O(h)$
corrections in the numerator of Eq.(\ref{eh}).
It is easy to check that the field corrections to the four-particle
states $|\rho>$ do not contribute up to the required $O(h)$ order,
so that it is enough to put in Eq.(\ref{eh}) the unperturbed state
$|\rho>$,  and to use
 the following expression for
 the perturbed ground state $|0_h>$
\begin{equation}
|0_h>=|0>-2h\sum_{\bf k}\frac{u_{\bf k}^2x_{\bf k}}{-2E_{\bf k}}|{\bf
k}> +O(h^2).
\end{equation}
Here $|{\bf k}>=\alpha_{\bf k}^{\dag}\beta_{\bf k}^{\dag}|0>$.
Then a straightforward calculation of the
 respective matrix elements yields
Eq.(\ref{mb}) for $\Delta m^{(2)}_b$.

\pagebreak

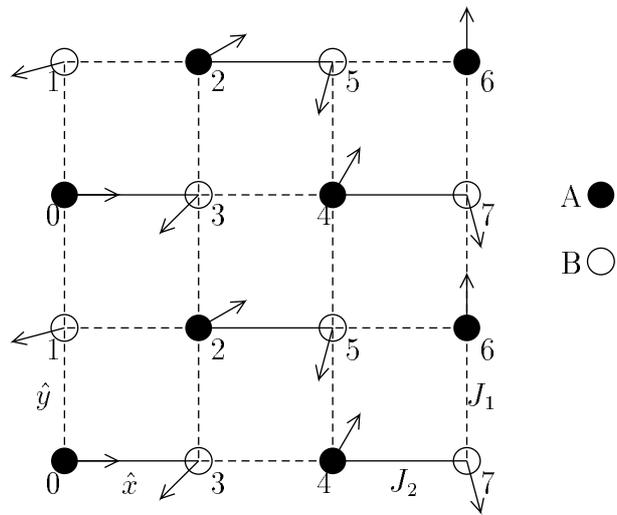
\begin{figure}
\caption{ The classical helicoidal state in a square-lattice
Heisenberg antiferromagnet with two kinds of regularly
distributed nearest-neighbor exchange bonds, $J_1$ and $J_2$.
The helicoidal phase is stable for $\alpha \equiv J_2/J_1 < -1/3$.
The spin orientations at $A$ and $B$ lattice sites are defined
by the angles $\theta_n=n\Phi$ and $\theta_n=n\Phi+\pi$,
respectively, where $n=0, 1, 2, ...$. The state is shown for
$\Phi=\pi/12$, i.e.,
$\alpha=-1/(1+ \protect{\sqrt{3}} )$ , and
$n=0, 1, .., 7$.}
\label{fig1}
\end{figure}

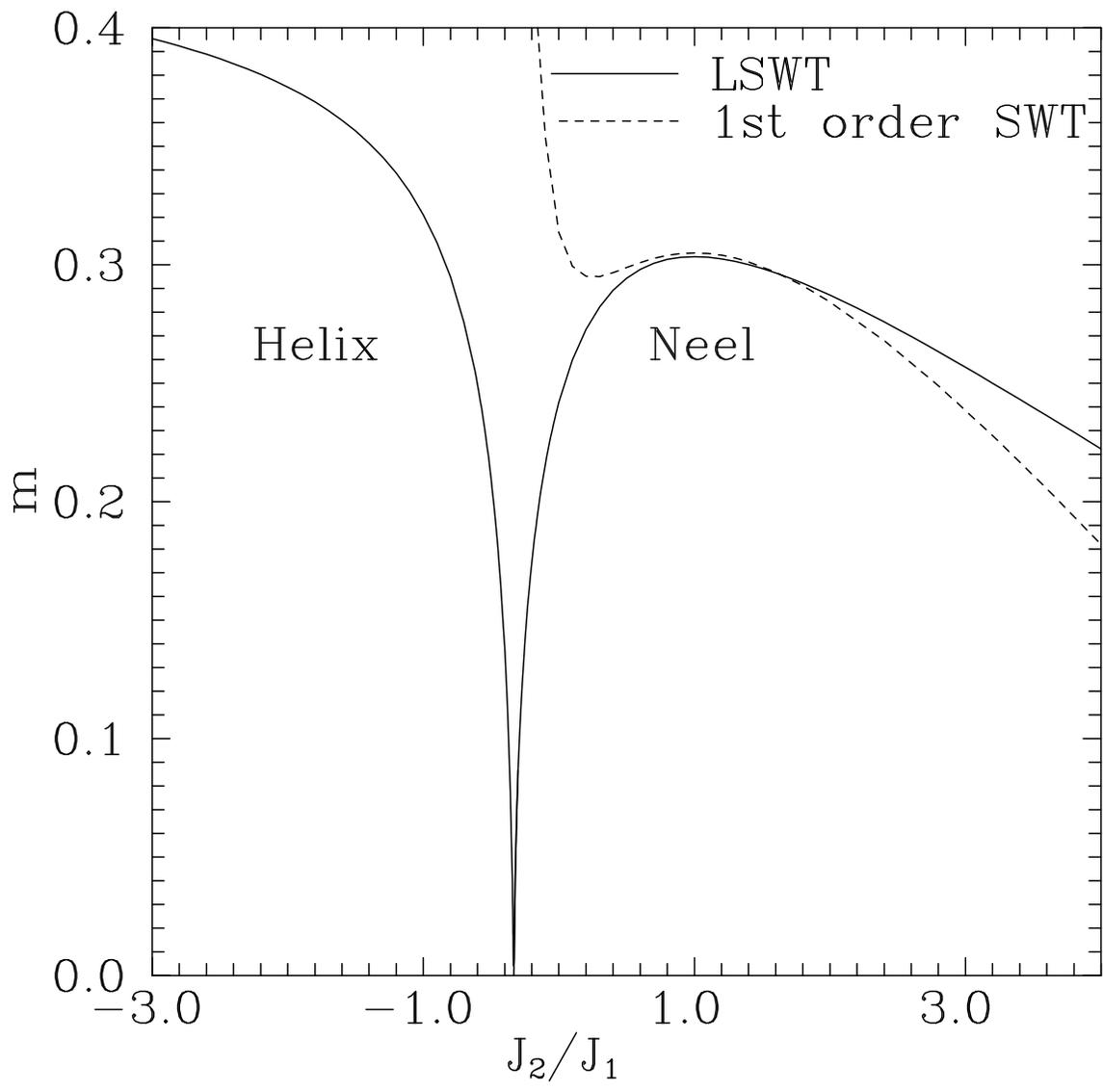
\begin{figure}
\caption{ Phase diagram of the $S=1/2$ system in a linear spin-wave
approximation (LSWT). The solid lines
 represent the on-site magnetization
$m$ as a function of the parameter
 $\alpha \equiv J_2/J_1$. The dashed
 line shows the diverging
first-order correction . There is a small region around
the classical transition point $\alpha_c=-1/3$
 ($ -0.336< \alpha <-0.331$)
where the magnetic states are melt. For $\alpha > \alpha_s$
a finite-gap quantum paramagnetic phase is stable.
$\alpha_s\approx 10.6$ in LSWT.}
\end{figure}

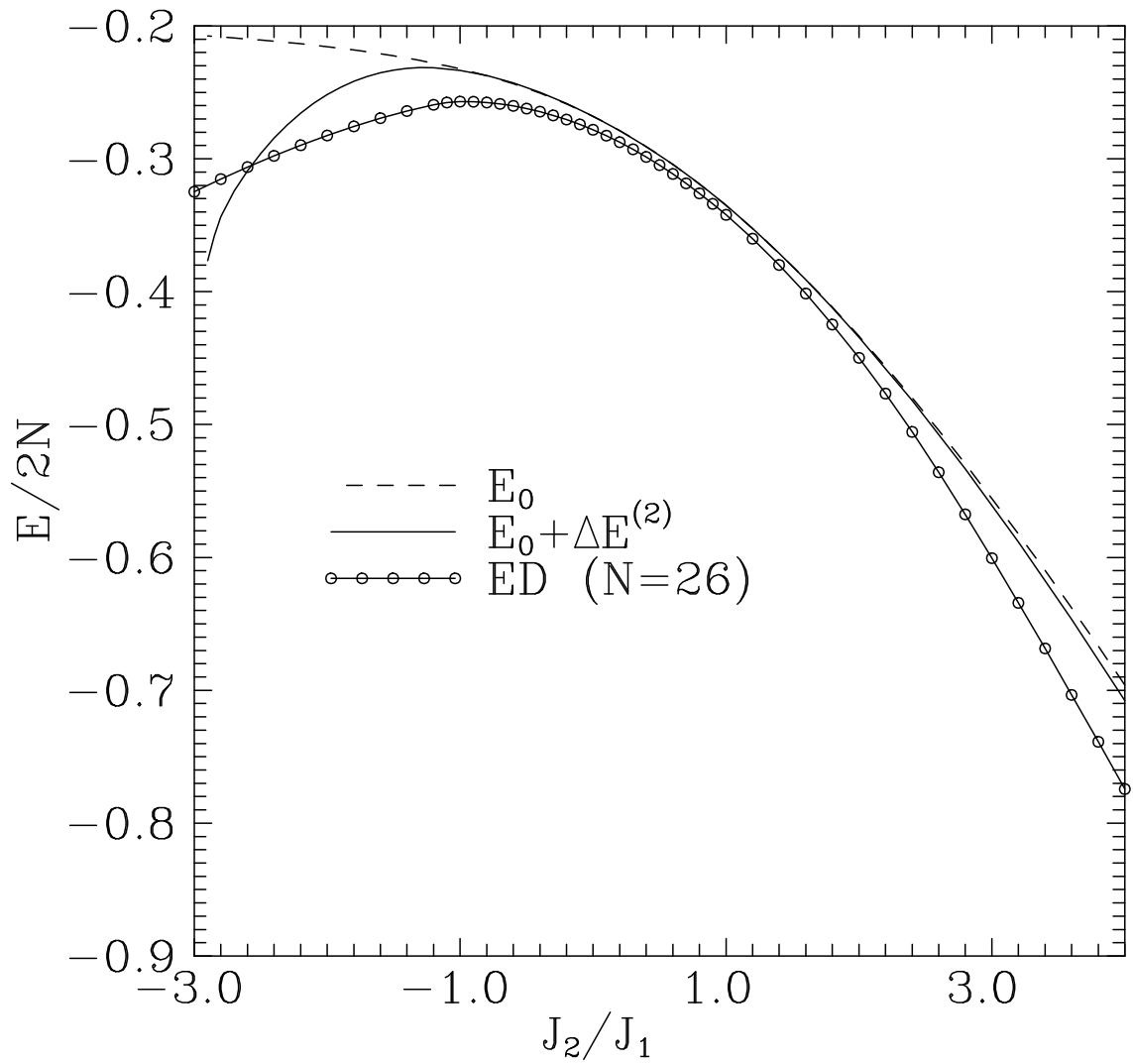
\begin{figure}
\caption{Ground-state energy of the model vs $\alpha
 \equiv J_2/J_1$, $S=1/2$. The open
circles are ED data for $N=26$. ED data for $N=16, 18, 20$ are
not shown but the deviations from the $N=26$ cluster are
very small. $E_0$ is the
zeroth-order mean-field result.   }
\end{figure}

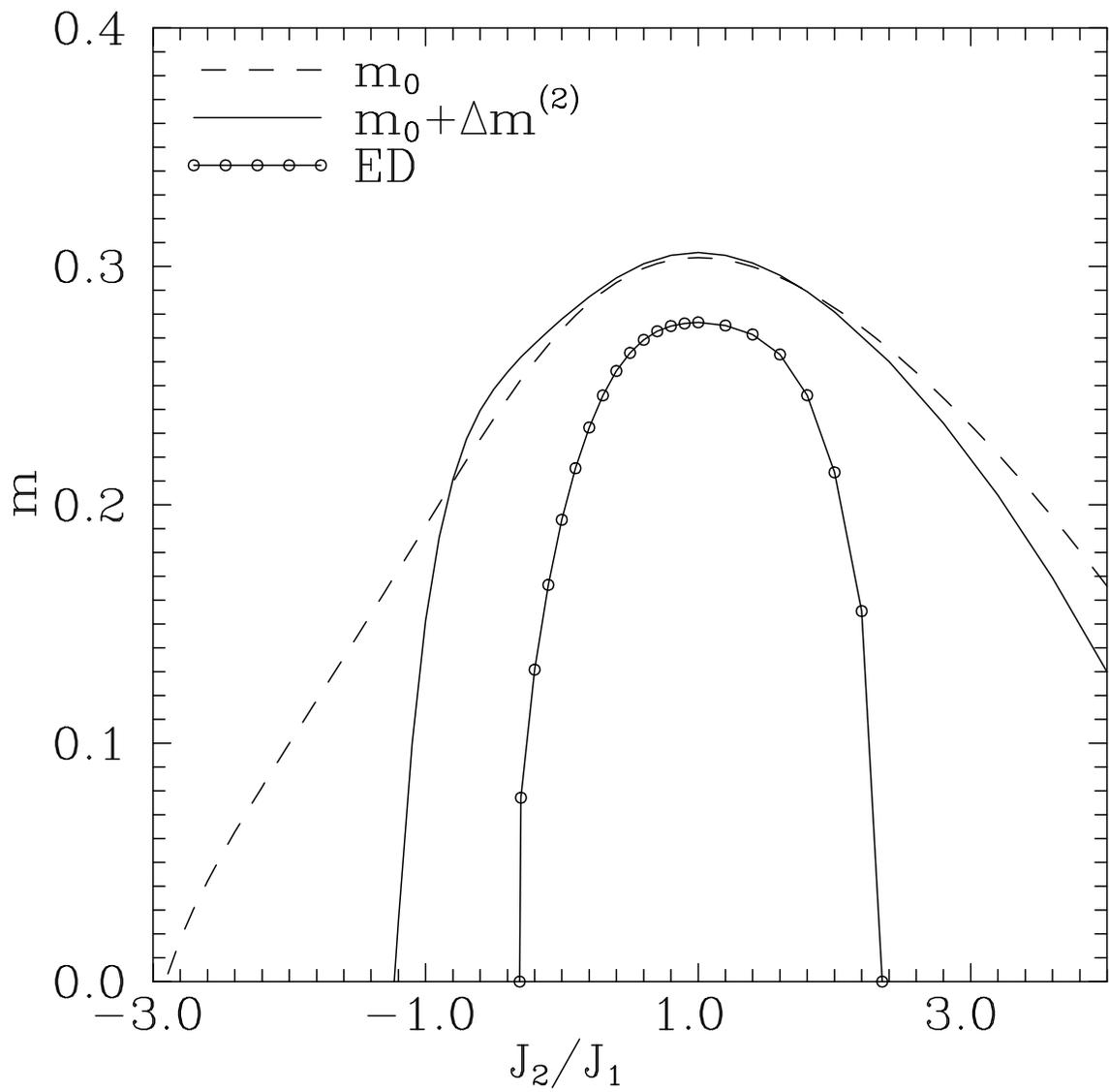
\begin{figure}
\caption{On-site magnetization in the N\'eel phase vs $\alpha \equiv
J_2/J_1$  calculated up to second order in $V_{DM}$, $S=1/2$.
 $m_0$ is the
zeroth-order mean-field result and
 $\Delta m^{(2)}$ is the second-order
$O(V_{DM}^2$) correction. The stars
 represent extrapolated ED data
for the on-site magnetization deduced from the antiferromagnetic
structure factor
\protect{ $M_s^2=\frac{1}{N^2}\sum_{i,j}(-1)^{i+j}<{\bf S_iS_j}>$}
 using
the cluster set $N=16, 18, 20, 26$ and the extrapolation
ansatz \protect{$M_s^2=M_s^2(\infty ) + constN^{-\frac{1}{2}}$} }
\end{figure}

\begin{table}
\caption{Data for the ground-state energy $E$
of the N\'eel phase calculated up
to  second-order corrections $\Delta E^{(2)}$ in respect to $V_{DM}$.
$\alpha=J_2/J_1$, $E=E_0+\Delta E^{(2)}$.
$E_0$ is the zeroth-order mean-field energy. $S=1/2$. }
\label{table1}
\begin{tabular}{cccccccc}
$\alpha$&$E_0/2N$&$\Delta E^{(2)}/2N$&$E/2N$&$\alpha$&$E_0/2N$&$\Delta
E^{(2)}/2N$&$E/2N$\\
\tableline
-2.0&  -0.21563&  -0.036  & -0.252  &
 0.4&  -0.29085&   0.00005& -0.29080\\
-1.8&  -0.21811&  -0.023  & -0.241  &
 0.6&  -0.30429&   0.00009& -0.30420\\
-1.6&  -0.22094&  -0.014  & -0.235  &
 0.8&  -0.31909&   0.00015& -0.31894\\
-1.4&  -0.22421&  -0.0078 & -0.2320 &
 1.0&  -0.33521&   0.00019& -0.33502\\
-1.3&  -0.22604&  -0.0054 & -0.2315 &
 1.2&  -0.35259&   0.00016& -0.35243\\
-1.2&  -0.22800&  -0.0035 & -0.2315 &
 1.4&  -0.37117&   0.00006& -0.37111\\
-1.1&  -0.23012&  -0.0021 & -0.2322 &
 1.6&  -0.39089&  -0.00015& -0.39104\\
-1.0&  -0.23241&  -0.0010 & -0.2334 &
 1.8&  -0.41168&  -0.00046& -0.41214\\
-0.8&  -0.23755&   0.0002 & -0.2374 &
  2.0&  -0.43348&  -0.00088& -0.43436\\
-0.6&  -0.24356&   0.0006 & -0.2430 &
  2.4&  -0.47987&  -0.00206& -0.48193\\
-0.4&  -0.25056&   0.0006 & -0.2500 &
 2.8&  -0.52964&  -0.00372& -0.53336\\
-0.2&  -0.25870&   0.00037& -0.25833&
  3.6&  -0.63800&  -0.00852& -0.64652\\
 0.0&  -0.26808&   0.00017& -0.26791&
  4.0&  -0.69606&  -0.01173& -0.70779\\
 0.2&  -0.27878&   0.00006& -0.27872
 \end{tabular}
 \end{table}

\begin{table}
\caption{Data for the on-site magnetization $m$
of the N\'eel phase calculated up
to second-order corrections $\Delta m^{(2)}$ in respect to $V_{DM}$.
$\alpha=J_2/J_1$, $m=m_0+\Delta m^{(2)}$.
$m_0$ is the zeroth-order on-site magnetization.  $S=1/2$.}
\label{table2}
\begin{tabular}{cccccccc}
$\alpha$&$m_0$&$\Delta m^{(2)}$&$m$&$
\alpha$&$m_0$&$\Delta m^{(2)}$&$m$\\
\tableline
-1.2 & 0.1726& -0.14   &0.03  &
  1.4 & 0.2996  &  0.0016& 0.3012    \\
-1.1 & 0.1819& -0.08   &0.10  &
  1.6 & 0.2952  &  0.0009& 0.2961    \\
-1.0 & 0.1911& -0.04   &0.15  &
  1.8 & 0.2894  & -0.0002& 0.2892    \\
-0.8 & 0.2095&  0.001  &0.211 &
 2.0 & 0.2824  & -0.0016& 0.2808    \\
-0.6 & 0.2274&  0.012  &0.239 &
  2.2 & 0.2743  & -0.0034& 0.2709    \\
-0.4 & 0.2444&  0.0112 &0.2556&
  2.4 & 0.2652  & -0.0054& 0.2598    \\
-0.2 & 0.2599&  0.0075 &0.2674&
  2.6 & 0.2553  & -0.008 & 0.247     \\
 0.0 & 0.2734&  0.0044 &0.2778&
  2.8 & 0.2446  & -0.010 & 0.235     \\
 0.2 & 0.2846&  0.0026 &0.2872&
  3.0 & 0.2331  & -0.014 & 0.219     \\
 0.4 & 0.2931&  0.0019 &0.2950&
  3.2 & 0.2210  & -0.017 & 0.204     \\
 0.6 & 0.2990&  0.0019 &0.3009&
  3.4 & 0.2081  & -0.021 & 0.187     \\
 0.8 & 0.3024&  0.0021 &0.3045&
  3.6 & 0.1946  & -0.025 & 0.170     \\
 1.0 & 0.3034&  0.0022 &0.3056&
   3.8 & 0.1805  & -0.030 & 0.150     \\
 1.2 & 0.3024&  0.0020 &0.3044&
  4.0 & 0.1657  & -0.036 & 0.130     \\
\end{tabular}
\end{table}


\begin{references}
\bibitem{ivanov2}
Notice that negative second-order corrections $\Delta E^{(2)}$ are
possible only for non-Hermitian operators, such as the Dyson-Maleev
interaction $V_{DM}$. Because the
mean-field energy $E_0$ can also be regarded as a
variational energy,
one can speculate that the observed
change of the sign indicates a kind of instability in the series.
\bibitem{bernu}
We have also tried to extrapolate with quantum normalized
$M_s^2$, as it has recently been suggested
by B. Bernu et al, Phys. Rev. B {\bf 50}, 10084 (1994),
 but the latter appoach predicts for $\alpha=1$
$m\approx 0.37$, which is less realistic.
\end{references}
\end{document}